\documentstyle[aps,manuscript]{revtex}

\newcommand{\be}{\begin{equation}}
\newcommand{\ee}{\end{equation}}

\def\fun#1#2{\lower3.6pt\vbox{\baselineskip0pt\lineskip.9pt
\ialign{$\mathsurround=0pt#1\hfil##\hfil$\crcr#2\crcr\sim\crcr}}}
\begin{document}

\def\baselinestretch{1.5}
\normalsize

\title{Comment on "Remark on the external-field method\\
 in QCD sum rules".}
\author{B. L. Ioffe}
\address{Institute of Theoretical
and Experimental Physics,
 B.Cheremushkinskaya 25, \\
 Moscow 117259, Russia}

 \maketitle
\begin{abstract}
It is proved, that suggested by Jin \cite{1} modified formalism in the
external-field method in QCD sum rules {\it exactly} coincides with
the formalism used before. Therefore, unlike the claims of ref.1, this
formalism cannot improve the predictability and reliability of
external-field sum rule calculations in comparison with those, done by the
standard approach.

PACS number(s): 12.38.Lg, 11.55.Hx

\end{abstract}

\def\baselinestretch{1.5}
\normalsize
\vspace{2mm}
\vspace{2mm}
In QCD sum rule calculations of hadronic properties in constant external
field the following problem arises. In the phenomenological part of the sum
rule besides the goal of the calculation -- the contribution of the lowest
hadronic state -- and the terms, corresponding to the transitions among
excited states, there appear the contributions of transitions from the
lowest to excited states. Unlike contributions of the transitions among
excited states, the latter are not exponentially suppressed by the Borel
transformation in comparison with the ground state term. In this aspect
there is an essential difference in the QCD sum rule calculations of
hadronic properties in the constant external fields -- the vertices -- and
the ones of the hadronic masses -- the polarization operators, -- where
excited states contributions are exponentially suppressed by the Borel
transformation. For this reason in the mass sum rules, it was possible
\cite{2} to use the rough model of hadronic spectrum -- the pole plus
continuum -- and the results were not too much sensitive to the model. In
the case of the calculations of hadronic matrix elements in the constant
external field the similar model can be used for the exponentially
suppressed contributions of transitions among excited states. But the terms,
which correspond to the transitions from the ground to the excited states,
should be considered exactly. The method, how to get rid of these terms,
which are a background in the sum rules, was suggested in the first
calculations of hadronic properties in the external fields -- the
calculations of nucleon magnetic moments \cite{3} (see also \cite{4}). The
idea was to exploit the different preexponential Borel parameters $M^2$
dependence of the ground and the background terms. By applying some
differential operator in $M^2$ to the sume rule it was possible to kill the
background term and to obtain the sum rules, where the excited states
contributions were exponentially suppresed, like in the QCD sum rules for
hadronic mass determination. Of course, the procedure of differention
deteriorates the accuracy of sum rule and this was a drawback of the method.

In the recent paper \cite{1} it was claimed, that a modified formalism is
invented, free from the mentioned above drawback, where  the excited states
contributions are exponentially suppressed relative to the ground state term
and this formalism has a potential to improve the predictability and
reliability of external-field sum rule calculation in comparison with the
method used before \cite{3,4}.

In  this comments I show, that the results obtained by the method, suggested
in \cite{1}  are identical to that used before and the method of \cite{1}
differs from \cite{3,4} only by transposition of intermediate mathematical
operations. So, no improvement of external-field sum rule calculations can
be achieved in this way.\\
The general expression of the QCD sum rule for three point vertex in a
constant external field (the linear response in the field in the correlation
function in terms of \cite{1})  has the form (see ref.5, eq.13):

$$\frac{\lambda^2 G}{(p^2 - m^2)^2} + \int\limits^{\infty}_{W^2}~ds~b(s)
\frac{1}{(s - m^2)}~\Biggl [ \frac{1}{(p^2 - m^2)} + \frac{\alpha(s)}{s -
p^2}\Biggr ] + \int\limits^{\infty}_{W^2}~ds \frac{\rho(s)}{(s-p^2)^2} =$$

\be
= \int\limits^{\infty}_0 ~ds \frac{\rho(s)}{(s - p^2)^2} + \sum_n ~c_n \frac
{1}{(p^2)^n}
\ee
The first term in the l.h.s. of (1) -- the phenomenological side of the sum
rule -- gives the contribution
of the ground state $h$. Here $G = <h \mid J \mid h > $ is the matrix
element over the state $h$, of the current $J$, interacting with the
external field, which we would like to find, $\lambda = < 0\mid \eta \mid h
>$, where $\eta$ is the quark current with the quantum numbers of hadron
$h$, $m$  is the hadron $h$ mass . The last term in the l.h.s. of (1)
corresponds to the transitions among excited states. Here the standard model
of continuum was used, where the continuum contribution is given by the
bare loop diagram and in the dispersion relation representation its
imaginary part starts from some threshold $W^2$. (The same expression for
continuum was used in \cite{1}). The first term in the r.h.s. -- the QCD
side of the sum rule -- gives the bare loop contribution. Here, as well as
in the continuum contribution in the l.h.s., the single variable dispersion
relation is assumed. (The use of a more general (see \cite{5}) double
dispersion relation does not influence our results.)  The last term in the
r.h.s. of (1) represents the higher order terms in the operator product
expansion. The second term in the l.h.s. is the background term discussed
above, corresponding to the transitions from the lowest to excited states.
The subtraction terms in the double dispersion relation, if it must be used,
are also accounted here. (The background term in (1) is more general, than
that, used in \cite{1}; the latter corresponds to $\alpha(s) = 1$).

Let us first treat (1) according to the method, proposed in \cite{3,4} and
perform first the Borel transformation. We have

$$ \frac{\lambda^2 G~ e^{-m^2/M^2}}{M^2} - \int\limits^{\infty}_{W^2}~ds
\frac{b(s)}{s - m^2}~\Biggl [ e^{-m^2/M^2} - \alpha (s)e^{-s/M^2} \Biggr ]
=$$

\be
= \int\limits^{\infty}_{W^2}~\rho(s)\frac{1}{M^2} e^{-s/M^2} + \sum_n ~c_n
\frac{(-1)^n}{(M^2)^{n-1}(n - 1)!},
\ee
where $M^2$  is the Borel parameter. In order to kill the nonsuppresed
exponentially background term, multiply (2) by $e^{m^2/M^2}$ and
differentiate over $1/M^2$. We get

$$\lambda^2 G - \int\limits^{\infty}_{W^2}~ds ~b(s)\alpha(s)~e^{-(s-m^2)/M^2}
= \int\limits^{\infty}_{W^2}~\rho(s)\Biggl ( 1 - \frac{s-m^2}{M^2}\Biggr )
e^{-(s-m^2)/M^2} ds +$$

\be
+ \sum_n~c_n \frac{(-1)^n}{(n-1)!}~\Biggl [ (n-1) + \frac{m^2}{M^2}\Biggr ]~
\frac{e^{m^2/M^2}}{(M^2)^{n-2}}
\ee
In (3) the contribution of transitions from the ground to excited states --
the second term in the l.h.s. -- are exponentially suppressed at least by
the factor $exp~[- (W^2 - m^2)/M^2].$

Instead of using this method Jin \cite{1}  proposed first multiply (1) by
$(p^2 - m^2)$ and then perform the Borel transformation. Going in this way
we have

$$ -\lambda^2 G e^{-m^2/M^2} + \int\limits^{\infty}_{w^2}~ds
~b(s)\alpha(s)e^{-s/M^2} = -\int\limits ~\rho(s)\Biggl ( 1 -
\frac{s-m^2}{M^2}\Biggr )e^{-s/M^2}ds -$$

\be
- \sum_n c_n \frac{(-1)^n}{(M^2)^{n-2}} ~\Biggl [ n - 1 +
\frac{m^2}{M^2}\Biggr ]
\ee
After multiplying by $- e^{m^2/M^2}$ the sum rule (4)  exactly, term by
term, concide with (3). Since in both approaches the final sum rules are
identical, no new results can be obtained by suggested by Jin \cite{1}
modification.\\
This investigation was supported in part by CRDF Grant RP2-132.

\end{document}